 \documentclass[preprint]{acmart}
\usepackage{subcaption}
\usepackage{booktabs}   
\usepackage{caption}    
\usepackage{tabularx}   
\usepackage{multirow}   
\usepackage[normalem]{ulem}
\usepackage[usenames,dvipsnames]{xcolor}
\AtBeginDocument{%
  }

\setcopyright{acmlicensed}
\copyrightyear{2018}
\acmYear{2018}
\acmDOI{XXXXXXX.XXXXXXX}

\acmConference[Conference acronym 'XX]{Make sure to enter the correct
  conference title from your rights confirmation emai}{June 03--05,
  2018}{Woodstock, NY}
\acmISBN{978-1-4503-XXXX-X/18/06}

\begin{document}

\title{Embedded vs. Situated: An Evaluation of AR Facial Training Feedback}

\author{Avinash Ajit Nargund}
\email{anargund@ucsb.edu}
\orcid{0009-0005-4224-3240}
\affiliation{%
  \department{Electrical and Computer Engineering}
  \institution{University of California, Santa Barbara}
  \city{Santa Barbara}
  \state{CA}
  \country{USA}
}

\author{Andrea M. Park}
\email{andrea.park@ucsf.edu}
\orcid{0000-0003-2761-2587}
\affiliation{%
  \institution{University of California, San Francisco}%
  \city{San Francisco}
  \country{USA}}

\author{Tobias Höllerer}
\email{holl@cs.ucsb.edu}
\orcid{0000-0002-6240-0291}
\affiliation{%
  \department{Computer Science}
  \institution{University of California, Santa Barbara}
  \city{Santa Barbara}
  \state{CA}
  \country{USA}
}

\author{Misha Sra}
\email{sra@cs.ucsb.edu}
\orcid{0000-0001-8154-8518}
\affiliation{%
  \department{Computer Science}
  \institution{University of California, Santa Barbara}
  \city{Santa Barbara}
  \state{CA}
  \country{USA}
}

\renewcommand{\shortauthors}{Nargund et al.}
\newcommand{\del}[1]{\iffalse#1\fi}
\newcommand{\add}[1]{#1}

\newcommand{\himg}[2][blue]{#2}
\setlength{\fboxrule}{1.5pt}
\begin{abstract}
While augmented reality (AR) research demonstrates benefits of embedded visualizations for gross motor training, its applicability to facial exercises remains under-explored. Providing effective real-time feedback for facial muscle training presents unique design challenges, given the complexity of facial musculature. We developed three AR feedback approaches varying in spatial relationship to the user: situated (screen-fixed), proxy-embedded (on a mannequin), and fully embedded (overlaid on the user's face). In a within-subjects study (N=24), we measured exercise accuracy, cognitive load, and user preference during facial training tasks. The embedded feedback reduced cognitive load and received higher preference ratings, while the situated feedback enabled more precise corrections and higher accuracy. Qualitative analysis revealed a key design tension: embedded feedback improved experience but created self-consciousness and interpretive difficulty. We distill these insights into design considerations addressing the trade-offs for facial training systems, with implications for rehabilitation, performance training, and motor skill acquisition.
\end{abstract}


\begin{teaserfigure}
  \includegraphics[width=\textwidth]{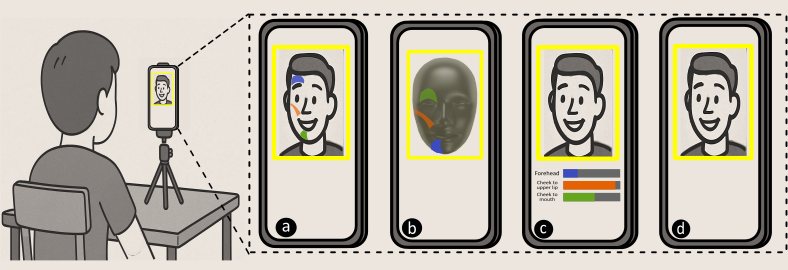}
  \caption{A participant engages in AR-guided facial exercises, receiving real-time feedback designed to improve muscle movement patterns. We evaluate three AR-based feedback conditions, (a) embedded \textit{ARSelfie} with feedback overlaid on the user's own face, (b) proxy-embedded \textit{Mannequin} with feedback on a 3D avatar, (c) situated \textit{BarChart} with feedback conveyed through bars located below the user's face and (d) a no-feedback \textit{Baseline} condition where the user can see themselves. We assessed how the spatial placement of feedback affects user performance, experience and cognitive load. Portions of this image were generated using AI.}
  \Description{}
  \label{fig:teaser}
\end{teaserfigure}

\received{20 February 2007}
\received[revised]{12 March 2009}
\received[accepted]{5 June 2009}

\maketitle

\section{Introduction}
Facial muscles play a crucial role in human communication, emotional expression, and various physiological functions~\cite{shackelford1997facial}. 
\add{Monitoring and training of facial muscle movements can offer benefits in specialized applications. These include actors and performers seeking to refine emotional expressivity~\cite{bloch1993alba, berry_dynamic_2022, ekman1997face}, public speakers aiming to enhance vocal~\cite{chen_utilizing_2015, fuyuno_multimodal_2018}, and patients undergoing rehabilitation for facial paralysis~\cite{barrios_dellolio_farapy_2021} or hypomimia~\cite{cai_break_2025}. In these settings, fine-grained guidance is essential to help users isolate and activate specific facial muscles. This work evaluates visualization efficacy independent of specific application. We believe that it is important across application domains to understand visualization impact on 1) exercise accuracy, 2) cognitive and task load, and 3) user preference.} 

\del{
While everyday interactions rely on casual observation of facial expressions, more precise monitoring and training of facial muscle movements can offer benefits in specialized domains such as refining emotional expressivity in performing arts~\cite{bloch1993alba, berry_dynamic_2022, ekman1997face}, improving delivery in public speaking~\cite{chen_utilizing_2015, fuyuno_multimodal_2018}, and supporting rehabilitation for conditions such as facial paralysis~\cite{barrios_dellolio_farapy_2021} and hypomimia~\cite{cai_break_2025}. These applications require fine-grained feedback to help users learn to isolate and selectively activate specific facial muscles.
}

Existing systems for facial muscle training provide limited forms of visual feedback, typically displayed adjacent to the user's face. For example, Farapy~\cite{barrios_dellolio_farapy_2021} overlays bar charts below the user's mirrored face to indicate muscle activation, while GY MEDIC~\cite{guanoluisa_gy_2019} and related systems place raw symmetry scores or text labels beside the face \cite{cai_break_2025}. These screen-adjacent visualizations are spatially disconnected from the region of movement, which may hinder a user's ability to interpret and act on the feedback~\cite{barrios_dellolio_farapy_2021}. While these tools support basic self-monitoring, they fall short of offering spatially integrated guidance that aligns with the movement itself.

In contrast, interactive systems for full-body and gross motor training have extensively leveraged spatially embedded feedback using augmented reality (AR)~\cite{sousa_sleevear_2016, zhu2022musclerehab, anderson2013youmove, physio@home}. These systems commonly align visual cues with the user's body or movement path, using AR mirrors, wearable displays, or head-mounted overlays~\cite{physio@home, sodhi_lightguide_2012, yu_persival_2024}. Such embedded visualizations~\cite{willett2016embedded} enhance user comprehension by co-locating feedback with the physical action. Though well-studied for limb and posture training~\cite{sigrist2013augmented, willett2016embedded}, these techniques remain underexplored for fine-grained and subtle movements of the face.

To address this gap, we investigate how the spatial placement of visual feedback affects user experience and performance during facial muscle training using mobile AR. We evaluate three visualization strategies to convey facial muscle activation that differ in their spatial relationship to the user’s face (see Figure~\ref{fig:teaser}): (a) screen-anchored \textit{BarChart},  (b) proxy-embedded \textit{Mannequin}, and (c) embedded \textit{ARSelfie} view. Inspired by the visualization used in Farapy~\cite{barrios_dellolio_farapy_2021}, \textit{BarChart} conveys muscle activations through proportional bar-fills placed at the bottom of the screen. In the \textit{Mannequin} paradigm, feedback cues are projected on a plain textured 3D model of the user's face while in the \textit{ARSelfie} view the cues are overlaid directly on the user’s face. These techniques represent distinct approaches along the ``WHERE''-axis of the design space of visual corrective feedback for XR motion guidance systems~\cite{yu_design_2024}. 

We conducted a within‐subjects study ($N=24$) in which participants performed three standardized facial exercises using the three visualization conditions and a no-feedback \textit{Baseline} condition where participants see their face in the camera without any feedback visualization. Our study was guided by the following research questions: 
\begin{enumerate}
    \item \textbf{RQ1:} Does \del{spatially-registered} \add{embedded} feedback (\textit{ARSelfie} and \textit{Mannequin} conditions) improve (a) exercise accuracy defined as the proportion of muscles activated correctly per repetition, and (b) rate of correctly executed exercise repetitions, when compared with the \del{screen-registered (} \add{situated} \textit{BarChart} condition\del{) feedback} or the \textit{Baseline} condition?
    \item \textbf{RQ2:} What are the cognitive and task loads associated with each feedback modality when compared to the \textit{Baseline} condition?
    \item \textbf{RQ3:} Which condition do users prefer?
\end{enumerate}

Our work makes the following contributions:
\begin{itemize}
\setlength\itemsep{.01pt}
    \item \del{AR feedback techniques for facial training.}We present three visualization approaches: situated, proxy-embedded, and fully embedded, that extend prior AR feedback work from gross motor to fine-grained facial muscle exercises.

   \add{
   \item From a within-subjects study ($N=24$), we  provide empirical evidence that embedded feedback which has been shown to be effective for full-body and gross motor training is also applicable for facial muscle training and results in better user experience compared to situated feedback which promotes precise activation of the facial muscles. 
   \item We synthesize the quantitative and qualitative results to propose design guidelines that inform how future systems can effectively balance exercise performance accuracy, feedback interpretability, and user comfort.
   }
\end{itemize}

\section{Related Work}

Our work is related to the use of AR, specifically situated and embedded visualizations in the context of instruction and feedback for facial motor learning. While several works have investigated the AR-based feedback for applications such as gait rehabilitation \cite{sekhavat2018gait}, upper limb training \cite{sousa_sleevear_2016} or general physical therapy~\cite{diller_skillar_2025}, their effectiveness and suitability for facial exercises remains largely unexplored in prior work. Only a few studies such as Farapy~\cite{barrios_dellolio_farapy_2021}, GY MEDIC~\cite{guanoluisa_gy_2019}, and ~\citet{cai_break_2025} have explored the use of AR-based feedback for facial exercises. In this section, we provide a brief overview of prior work to contextualize the contributions of our research.  

\subsection{Instruction and Feedback for Motor Learning}
Acquiring a motor skill is a complex process that depends on effective instruction and feedback. Motor learning involves three sequential stages: cognitive, associative and autonomous~\cite{fitts1967human}. Feedback is most critical during the initial cognitive stage, where the learner is forming a mental model of the movement, and in the associative stage, where they are refining their technique. While learners receive intrinsic feedback through their own sensory systems (e.g., proprioception), augmented feedback provided by an external source, such as a human coach or a computer, is often helpful in accelerating learning and overcoming performance plateaus~\cite{schmidt2008motor}.

One of ways to categorize augmented feedback is by distinguishing between \textit{Knowledge of Results (KR)} which informs the learner about the result of their action (e.g., did they hit the target or not) reducing the \textit{gulf of evaluation}~\cite{norman1986cognitive}, and \textit{Knowledge of Performance (KP)}, which provides information about the correctness and characteristics of the movement itself (e.g., hand was not raised fast enough) which helps diminish the \textit{gulf of execution}~\cite{norman1986cognitive}. AR-based training systems are capable of sensing user movements in real-time and delivering detailed KP that would be challenging for a human tutor to convey accurately~\cite{sigrist2013augmented}. 

Our work explores the use of embedded AR visualizations for providing precise, real-time KP for facial exercises. We compare this approach with situated AR visualizations to assess their impact on user performance, experience, and preference.     

\subsection{Feedback Visualization}
Situated visualization, introduced by White \cite{white2009interaction},  refers to the spatial placement of data visualizations close to the physical entities or referents it is associated with. While situated visualizations remain adjacent to the referents, embedded visualizations go further by aligning visuals with the physical extents of the referents~\cite{willett2016embedded}.
Research in feedback systems for motor skill acquisition and physical rehabilitation has explored a variety of visualization paradigms designed to bridge the gap between a user's current and target state. These systems primarily differ in how the feedback is spatially related to the user's body. The spatial relationship strongly affects the interpretability of the guidance and the cognitive load imposed on the users~\cite{sigrist2013augmented, willett2016embedded}.

Within the broad spectrum of visualization paradigms, prior work in AR-based training systems has used three main strategies: situated, proxy-embedded, and embedded.

\paragraph{Situated feedback}
Situated visualizations separate the feedback from the body, displaying it in the surrounding environment, such as projecting foot placements on a treadmill~\cite{sekhavat2018gait}, displaying life-size visuals of an expert to help users with rock climbing~\cite{kosmalla2017climbvis} or providing text-based exercise feedback~\cite{da2016mirrarbilitation, caserman2021full}. 

\paragraph{Proxy-embedded feedback}
Proxy-embedded visualizations map feedback onto avatars or skeleton models scaled to the user's size and body dimensions. YouMove~\cite{anderson2013youmove} visualized movement guidance on a skeleton displayed on a large AR mirror while MuscleRehab~\cite{zhu2022musclerehab} provided feedback by displaying muscle activation levels on a virtual 3D model. 

\paragraph{Embedded feedback}
Embedded feedback minimizes perception-action gaps by aligning cues directly with the body~\cite{willett2016embedded}. LightGuide~\cite{sodhi_lightguide_2012} projected 2D and 3D arrows directly onto user's hands to guide mid-air gestures, improving hand movement accuracy by nearly 85\% when compared to video-based guidance while SleeveAR~\cite{sousa_sleevear_2016} projected color-coded feedback onto a custom sleeve to support shoulder rehabilitation. Some systems use virtual mirrors to overlay visuals on the user's reflection. These systems combine the familiarity of mirrors with augmented feedback to either guide users ~\cite{physio@home} or highlight errors made by the user~\cite{trajkova2018takes}.

While the efficacy of these paradigms has been demonstrated for upper-limb rehab and gait training, their application to facial exercises, which involve fine motor movement, is underexplored. Existing facial exercise systems have predominantly adopted situated visualizations, displaying feedback adjacent to the user's face. For example, Farapy~\cite{barrios_dellolio_farapy_2021} and ~\citet{cai_break_2025} use AR overlays near the user's selfie view to provide performance feedback. However, the separation of the feedback from the face introduces a disconnect, with Farapy~\cite{barrios_dellolio_farapy_2021} participants reporting that bars at the bottom of the screen made interpreting feedback harder, and suggested that embedded cues, such as color changes directly on the facial muscles, would have been preferable.

In our work, we investigate the effectiveness of three visualization strategies explicitly designed for facial motor training. Using these visualization conditions, we first examine whether embedded visualization, which have demonstrated benefits for limb and whole-body training~\cite{yu_design_2024}, can be applied to facial exercise training. We then compare how the spatial placement of feedback relative to the face influences accuracy of facial movements, the ease of understanding guidance, and the cognitive load it imposes on the users.

\add{
\subsection{On-Face Visualization}
Prior research has utilized user's reflection to display personal health informatics~\cite{alhamid2012multi, ceccaroni2004magical, henriquezMirrorMirrorWall2017, riglingYourFaceVisualizing2023, subramonyam2015sigchi}. Wize Mirror~\cite{henriquezMirrorMirrorWall2017} used data from multiple sensors to estimate cardio-metabolic risk factors and display a composite ``Wellness Index'' next to the user's reflection. Similarly, ~\citet{riglingYourFaceVisualizing2023} explored the use of AR face filters to visualize fitness tracker metrics, such as step count and sleep time.}

\add{
Building on prior work that visualizes health data on the body or face to reduce the disconnect between data and its context, we extend this approach to facial-muscle training by visualizing muscle activations in real-time on the corresponding facial anatomy to provide users with immediate, contextually grounded feedback.
}

\begin{figure}[!t]
    \centering
    \himg{
    \includegraphics[width=\linewidth]{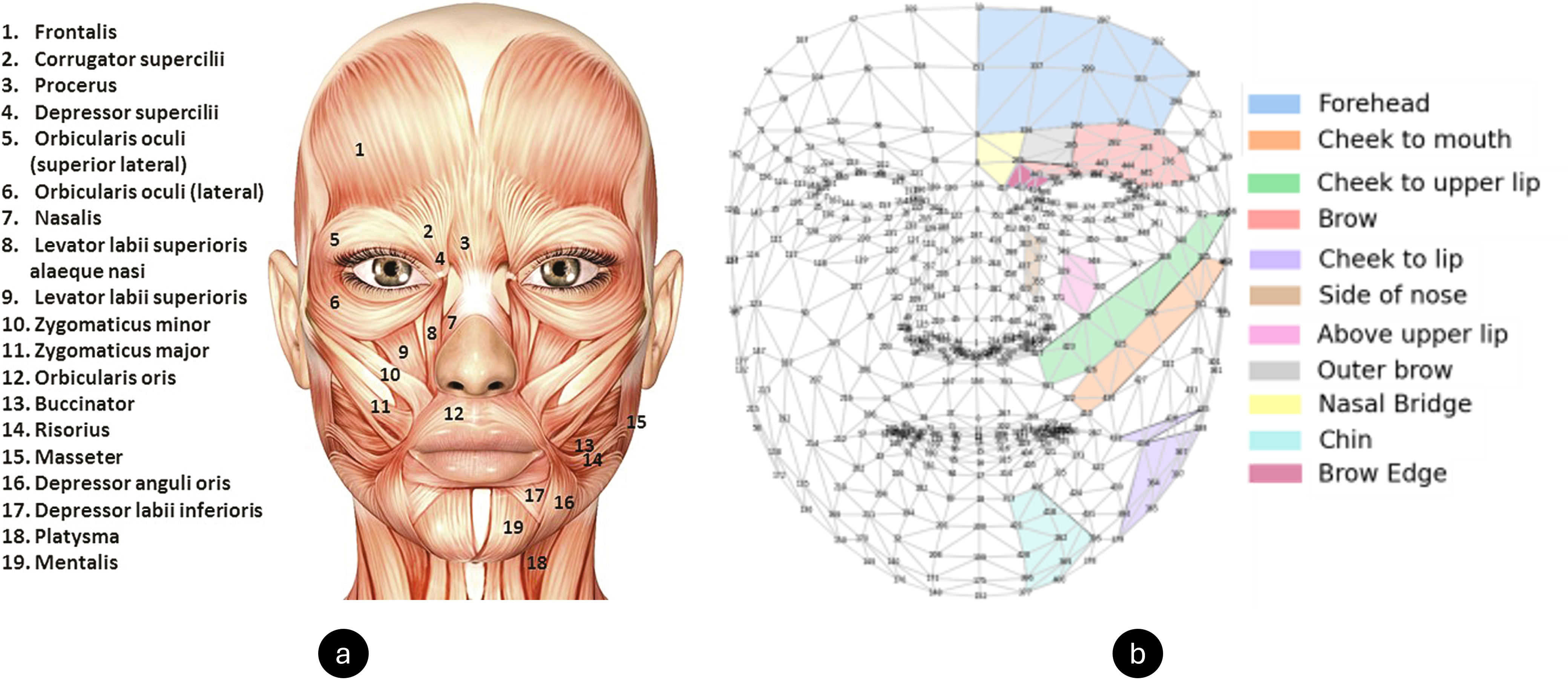}
    
    \caption{\add{The facial muscle anatomy chart (a) (from ~\cite{nestor2017}) that we used to map subsets of the 468 facial landmarks 3D face mesh (b) provided by Google ARCore \textit{AugmentedFaces} API to the target muscles in our study.}}
    \label{fig:face_mesh}
    }
\end{figure}
      
\section{Study Design and Implementation}\label{sec:design_impl}
To investigate our research questions, we developed a mobile AR application using the Unity\footnote{\texttt{https://unity.com/}} game engine. The application uses Google ARCore's \textit{AugmentedFaces} Unity API~\footnote{\texttt{https://developers.google.com/ar/develop/unity-arf/augmented-faces/developer-guide}} for facial tracking. This API provides a 468-point 3D face mesh (Figure~\ref{fig:face_mesh}b) and tracks the face in real-time using the front-facing camera of the phone. We leverage this dense face mesh to calculate muscle activation and to spatially register feedback overlays onto the user's face. The landmark-based approach offers two key advantages over using raw image pixels. First, the pre-trained machine learning (ML) models in ARCore are optimized for mobile inference and trained on large-scale datasets, making them robust to variations in lighting, head pose, and user appearance. Second, the 3D mesh provides a semantically organized and anatomically grounded representation of the face. This directly facilitates derivation of interpretable muscle activation estimates by measuring localized landmark displacements, whereas pixel-based methods often rely on indirect inference and require large, labeled datasets to handle variations in appearance, expression, and lighting. 

We design and implement four experimental visualization conditions : \textit{ARSelfie} (embedded), \textit{Mannequin} (proxy-embedded),  \textit{BarChart} (situated) and \textit{Baseline}. In this section, we describe our design rationale, system architecture, and features of each visualization condition.

\subsection{Mapping Landmarks to Muscles}
To provide muscle-specific feedback, we manually map subsets of the 468 ARCore facial landmarks to their corresponding facial muscles. We leverage the biomechanical principle that facial muscles, unlike most other muscles, are attached directly to the skin rather than bone~\cite{cattaneo_facial_2014}. This direct attachment means that muscle contraction causes localized skin deformation, enabling the use of facial landmark clusters \add{as a proxy} to estimate muscle activations~\cite{xu2022facial, shu2025facial}. \add{This method is similar to the approach in prior work by Hasebe et al. \cite{hasebe2024}, which establishes that landmark cluster movements correlate with integrated electromyography (iEMG) estimates of muscle activations, particularly for the mouth (significant correlation) and eye muscles.} By overlaying the ARCore face mesh onto facial musculature anatomy charts~\cite{marur_facial_2014}, we visually identified clusters of facial landmarks situated over particular muscles, whose displacement indicates activation. The accuracy of the mapping was reviewed and validated by our second author, who is a facial nerve surgeon.

\subsection{Participant Calibration}
In the calibration phase, the user is first prompted to hold a neutral facial expression, during which the system records the base position of all the facial landmarks. Next, the user performs each target exercise with maximum effort (e.g., smiling as wide as possible) and the system records the 3D landmark positions at peak activation. The maximum movement range of each muscle is then calculated as the 3D Euclidean distance between the centroid of its landmark cluster at peak activation and in the neutral expression. This calibration establishes a personalized reference scale for determining activation levels. 

\begin{figure}[t]
    \centering
    \includegraphics[width=\linewidth]{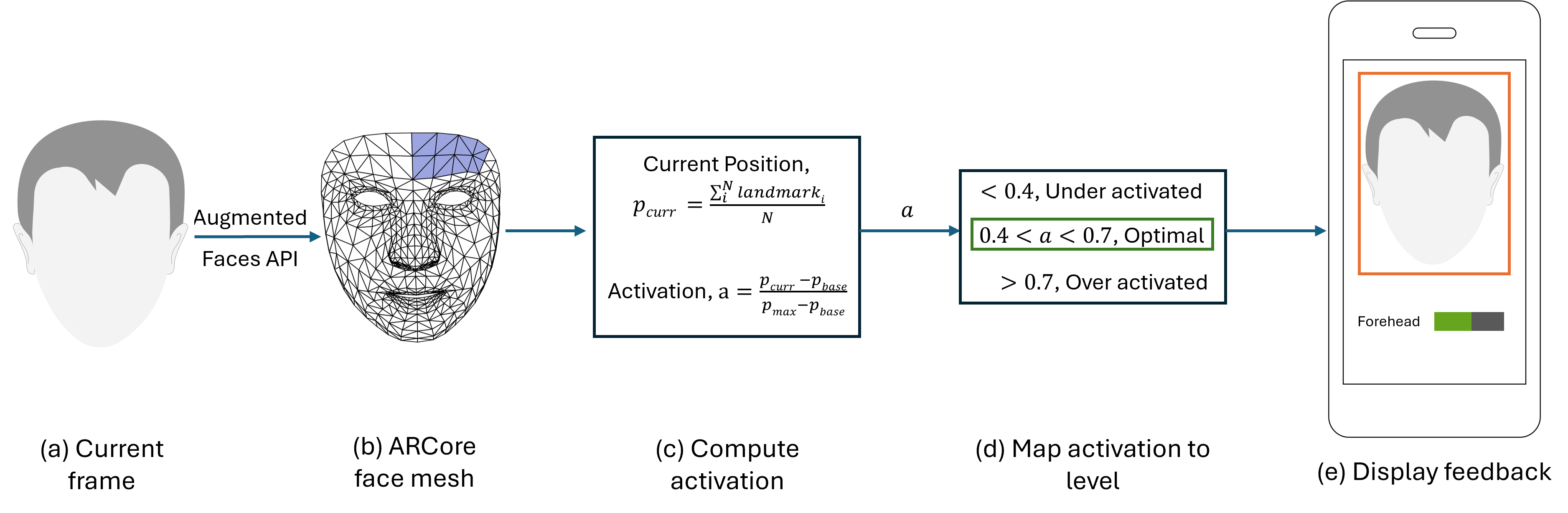}
    \caption{The activation estimation process, demonstrated with the left forehead muscle in the \textit{BarChart} condition. First, the phone's front camera captures the user's face and the \textit{AugmentedFaces} API returns a face mesh. The landmarks corresponding to a particular muscle are used to estimate its current position. This position along with the muscle's baseline and peak positions are used to compute a raw activation score. This score is mapped to an activation label and both are used to provide feedback to the user.}
    \label{fig:activation_estimation}
\end{figure}

\subsection{Muscle Activation}

During an exercise, the system estimates muscle activation by tracking the displacement of specific facial landmark clusters \add{(see Figure~\ref{fig:activation_estimation})}. The centroid of each cluster is used as the position of the corresponding muscle, and the system continuously measures displacement as the Euclidean distance between its current position and its initial position in a neutral expression. 
The muscle activation is then computed as the ratio of current displacement to maximum movement range established in the calibration step. The normalization accounts for inter-individual variability, as it creates a personalized 0 to 1 scale for all the muscles of each user based on their range of motion. The activation score is then mapped to a three-class label. Scores below $0.4$ are classified as under-activation, scores above $0.7$ as over-activation and scores in between as optimal. These thresholds were determined through pilot testing and consultation with a facial surgeon. The three-way classification is similar to the approach used in prior work~\cite{barrios_dellolio_farapy_2021}.

For the study, we use a landmark displacement-based heuristic instead of a machine learning (ML) model to ensure deterministic and stable muscle activation estimates. Controlling the variability at the activation estimation level enables us to isolate the effects of the visualization design and avoid potential confounds from noisy or inaccurate ML predictions. Additionally, this heuristic approach minimizes computation load on the mobile device, ensuring the feedback interface is fast and responsive. 

\begin{figure*}[b]
    \centering
    \himg{
    \includegraphics[width=\linewidth]{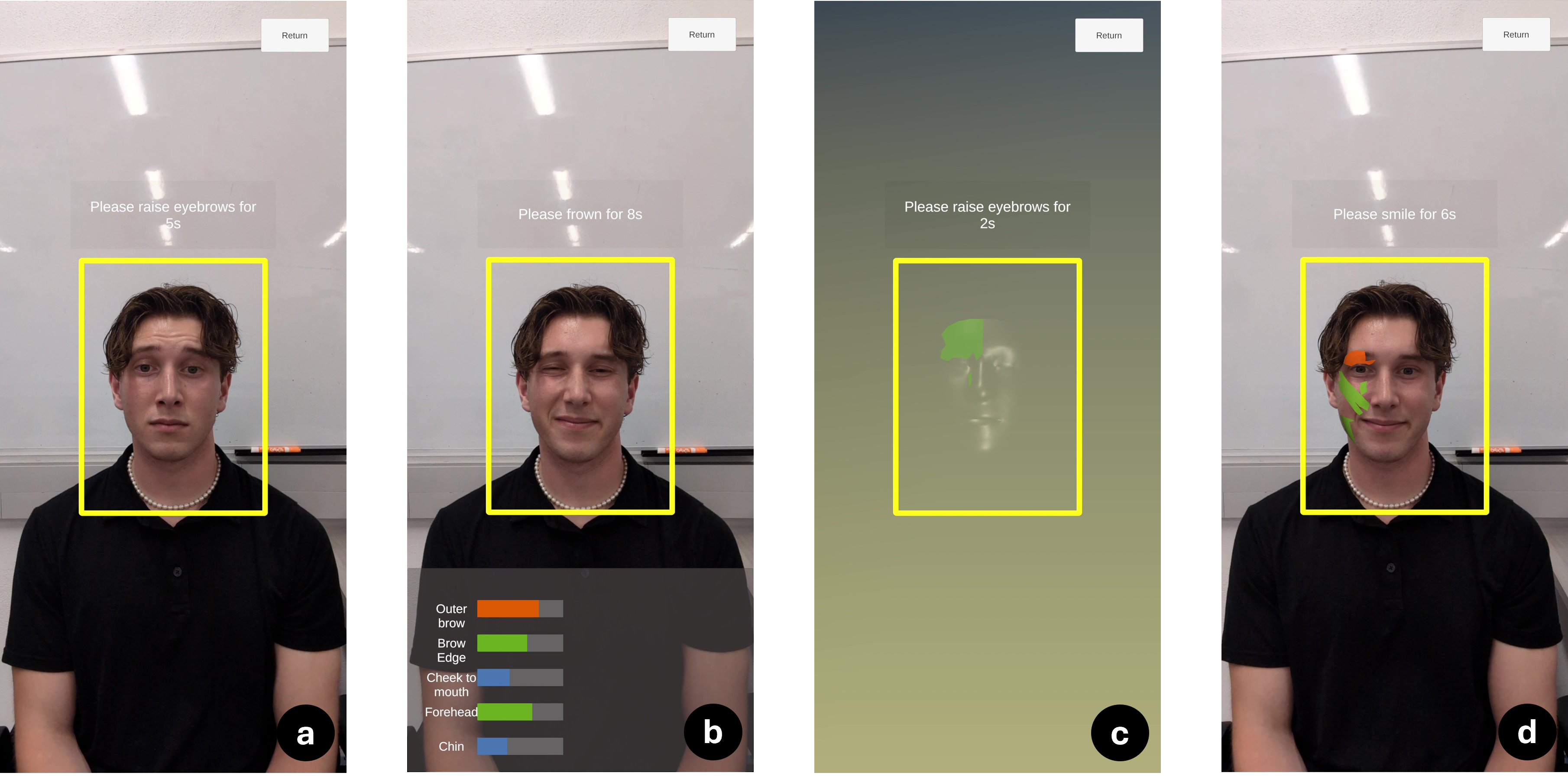}
    
    \caption{\add{The four experimental conditions used in the study, showing a participant performing the three facial exercises :  (a) \textit{Baseline} (\textit{Eyebrow Raise}), (b) \textit{BarChart} (\textit{Reverse Frown}), (c) \textit{Mannequin} (\textit{Eyebrow Raise}) and (d) \textit{ARSelfie} (\textit{Smile}).}}
    \label{fig:ui_with_exercises}
    }
\end{figure*}

\subsection{Visualization Conditions}
To explore how the spatial location and level of embodiment of the feedback impacts users, we designed four visualization conditions \add{(see Figure~\ref{fig:ui_with_exercises})}.

\paragraph{Baseline}
In the \textit{Baseline} condition, the user's live camera feed is displayed with no activation score-based feedback. 

\paragraph{Situated Visualization: BarChart}
The \textit{BarChart} condition provides muscle activation feedback through horizontal bar charts placed at the bottom of the device screen, detached from the user's live camera feed shown at the top. Each bar corresponds to a target muscle in the exercise and is labeled using the muscle's common name (e.g.,\textit{Frontalis} muscle is labeled ``Forehead''). The width of the bar fill is proportional to the muscle's raw activation, which provides continuous feedback on the activation intensity to the user. The bar fills are colored based on the activation level: blue for under-activated, green for optimally activated and orange for over-activated. 

\paragraph{Proxy-Embedded Visualization: Mannequin}
The \textit{Mannequin} condition provides visual feedback on a dynamic 3D facial avatar created using the ARCore face mesh. The avatar replicates the user's facial movements in real time via mesh deformation but omits photorealistic textures, appearing instead as a neutral proxy face. Feedback is conveyed through colored overlays aligned with specific muscle regions, generated by segmenting the mesh into sub-regions that correspond to the targeted muscles.  

\begin{figure}[t]
    \centering
    \includegraphics[width=\linewidth]{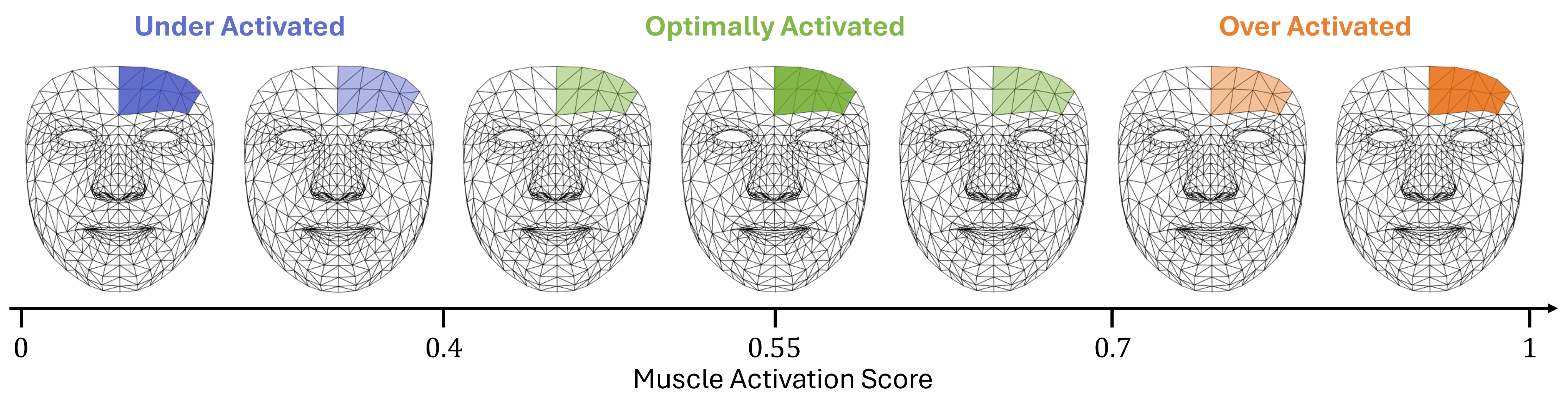}
    \caption{
    Opacity-modulated feedback in the embedded conditions, illustrated for the forehead muscle on the ARCore face mesh. Color indicates the activation zone: blue for under-activation, green for the optimal range, and orange for over-activation. Opacity encodes distance from the optimal range: the blue overlay fades as activation approaches green, the green overlay is most opaque at the midpoint and more transparent near its lower and upper thresholds, and the orange overlay increases in opacity as activation rises further beyond optimal.}
    \label{fig:op_modulation}
\end{figure}

This visualization condition uses both color and opacity to communicate feedback. \del{similar to the visualizations by Ikeda et al.~\cite{ikeda2018ar}} \add{This approach is consistent with the method used by Ikeda et al.~\cite{ikeda2018ar} and aligns with the ``Decal'' pattern for situated visualizations described by Lee et al.~\cite{lee_design_2023}} (see Figure~\ref{fig:op_modulation}). The color of an overlay represents the discrete activation level: blue for under-activated, green for optimal, and orange for over-activated. The opacity of the overlay is indicative of the absolute difference between the current and optimal activation score. In the under-activated range (scores below $0.4$), the blue overlay's opacity decreases as the activation approaches the optimal threshold, encouraging the user to activate their muscles further. In the optimal range ($0.4-0.7$), the green overlay is most opaque at the center of the range (a score of $\sim0.55$) and becomes gradually more transparent toward the lower and upper bounds of optimality. 
In the over-activated range (scores above $0.7$), the orange overlay’s opacity increases as the activation score increases, indicating excessive activation.

\paragraph{Embedded Visualization: ARSelfie. }
The \textit{ARSelfie} condition uses the same design, feedback mechanism and colors as the \textit{Mannequin}. The main difference is that the feedback is overlaid directly onto the user's face in the live video feed rather than on a proxy avatar face. 

\subsection{Data Logging}
The mobile application logs the raw activation scores for all target muscles and identifiers for the current exercise into separate files for each condition. This data is captured at the application's frame rate (synchronized to Unity's \texttt{Update()} cycle) and is used to compute performance metrics including the proportion of exercise time during which muscles were optimally activated, the time taken to achieve the first optimal repetition, and the total number of optimal repetitions within the exercise window.  

\section{User Evaluation}
\begin{figure}[!t]
    \centering
    \himg{
    \includegraphics[width=\linewidth]{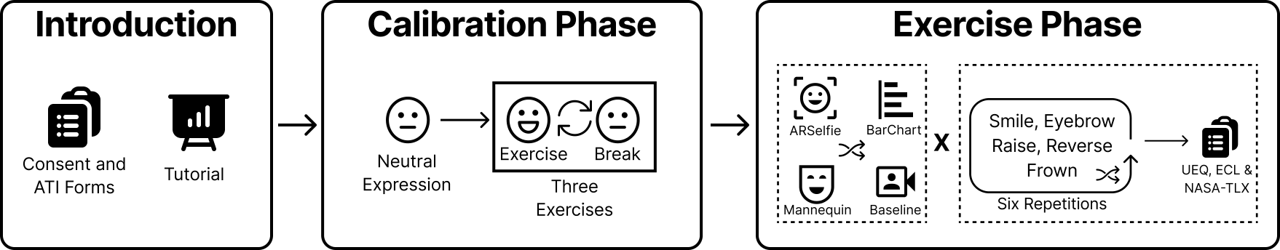}
    
    \caption{
    \add{
    Study procedure. After providing informed consent and completing the ATI questionnaire~\cite{franke2019personal}, participants received a tutorial on exercises, study task and feedback conditions through a presentation, after which they switched to the system for the calibration and exercise phases. In the calibration phase participants first maintained a neutral expression (10 seconds) and then performed the three exercises at maximum effort (10 seconds each with 5-second breaks). In the exercise phase, participants completed four counter-balanced trials (Latin-square design), one per condition. Each trial consisted of six repetitions of the exercise cycle (three exercises performed for 10 seconds in randomized order with 5-second breaks). After each trial, participants completed the NASA-TLX~\cite{hart_development_1988}, UEQ~\cite{schrepp2017construction}, and a custom extraneous cognitive load questionnaire.
    }
    }
    \label{fig:user_eval_steps}
    }
    
\end{figure}

We conducted a user study to examine how different spatial placements of visual feedback for facial motor training influence user performance,  perceived task load and experience as they perform facial exercises. To run the study, we developed a custom application that provided three types of AR-based feedback, \textit{ARSelfie}, \textit{Mannequin} and \textit{BarChart}, as well as a no-feedback \textit{Baseline} (described in Sec.~\ref{sec:design_impl}). \del{Three facial exercises were selected in consultation with a facial surgeon to ensure medical validity.} The study was conducted in a quiet, indoor room with consistent lighting. The application ran on a Google Pixel 8 Pro smartphone, placed $0.5$ meters in front of the participant. A chin rest was used to stabilize the participant's head and minimize face tracking inconsistencies from unintended head movements. The \add{facial exercises used in the study,} study procedure (outlined in Figure~\ref{fig:user_eval_steps}) and the task are described below. \del{and shown in Figure~\ref{fig:user_eval_steps}.}

\add{
\subsection{Facial Exercises}
Participants performed three facial exercises, smile, eyebrow raise, and reverse frown, each requiring the activation of muscles of different sizes from distinct facial regions (See Figure \ref{fig:ui_with_exercises}). The smile relied on the engagement of  muscles around the mouth. The eyebrow raise involved raising the eyebrows toward the hairline which is mainly controlled by the forehead muscle. The reverse frown, commonly known as "face scrunch," required participants to draw their eyebrows down and together while moving the muscles in their cheeks and chin upwards. This compound movement requires coordinating muscle activation across the upper and lower face.
} 

\subsection{Study Procedure}
Participants began the study by providing informed consent (study approved by our local IRB under protocol \#anonymous), demographic information and completing the Affinity for Technology Interaction (ATI) questionnaire~\cite{franke2019personal}, which assesses individual differences in interaction and engagement  with technology. Participants were briefed on the purpose of the study and introduced to the three facial exercises through a tutorial. The tutorial also explained the visualization encodings including the color scheme and opacity modulation used across conditions. \add{Participants also learned that optimal repetitions  indicated by muscle-feedback indicators turning green for all muscles for that exercise.} 

Following the tutorial, participants completed a calibration phase. In this phase, they were first instructed to maintain a neutral expression for 10 seconds to record their baseline facial landmark positions. Next, they performed the three exercises with maximal effort with a 5 second break between exercises. This calibration also helped familiarize participants with the exercises \add{they would perform in the exercise phase (detailed in Section~\ref{sec:study_task})}. 

During the exercise phase \del{(detailed in Section~\ref{sec:study_task})}, after each visualization condition, participants completed the NASA-TLX questionnaire~\cite{hart_development_1988}, the User Experience Questionnaire (UEQ)~\cite{laugwitz_construction_2008} and a custom  questionnaire adapted from~\cite{klepsch_development_2017} to assess extraneous cognitive load \add{(ECL)} (\add{Table}~\ref{tab:ecl_questions}). Measuring the extraneous cognitive load allows us to estimate how demanding each visualization condition made it for the participants to extract task-relevant information and perform the exercises effectively. At the end of the study, participants completed a post-study questionnaire where they ranked the conditions in order of preference and provided rationale for their preferences in a semi-structured interview with the researcher.

\subsection{Study Task}\label{sec:study_task}
\add{
During the exercise phase, participants completed four trials, one per visualization condition. Trial order was counterbalanced using a Balanced Latin Square to minimize learning effects and reduce potential carryover between conditions. Each trial consisted of six sets, and each set contained all three exercises with their order permuted to mitigate sequence effects. For every exercise, participants worked for 10 seconds followed by a 5-second rest. They were instructed to perform as many optimal repetitions as possible within each 10-second window, where a repetition was counted as optimal only when all muscle-feedback indicators appeared green. The exercise and break durations were determined through pilot testing to provide sufficient time for participants to perform multiple repetitions per exercise window while minimizing fatigue over the full study session.
}

\subsection{Participants}
Following initial pilot tests, a power analysis was conducted to determine the required sample size. Assuming small-to-medium effect size ($f = 0.2$), significance level of $\alpha = 0.05$, a statistical power of $1 - \beta = 0.8$ and correlation of $0.7$ among repeated measures, the analysis indicated a target sample of 22 participants. We subsequently recruited 24 participants (10 female, 14 male) aged between 18 to 30 (median = 20) years using internal mailing lists. \add{ All participants had normal or corrected-to-normal vision. Seventeen participants reported no color vision deficiencies. Two participants self-reported minor color vision deficiencies but confirmed that they could distinguish between colors used in the study. The remaining five participants did not disclose their color vision status but reported no difficulty distinguishing colors before or during the study.} All participants were compensated \$15 for their time.

\section{Quantitative Results}
\del{We analyzed the objective performance, user experience,  and task load data to assess the effect of the feedback conditions.}
\add{
We evaluated the following dependent variables to assess the effect of feedback conditions: (a) performance metrics including exercise accuracy, time to first optimal repetition and  number of repetitions, (b) user experience ratings measured through the User Experience Questionnaire (UEQ)~\cite{schrepp2017construction}, (c) task load measured via the NASA-TLX~\cite{hart_development_1988}, and (d) extraneous cognitive load (ECL) scores.
}
A Shapiro-Wilk test was conducted on all dependent variables to assess normality. We used a repeated measures ANOVA when the assumption of normality was met and a Friedman test otherwise. All post-hoc pairwise comparisons were performed using Bonferroni-Holm correction.   

\subsection{Exercise Performance}
\begin{figure}[!t]
    \centering
    
    \begin{subfigure}[b]{0.3\textwidth}
        \centering
        \includegraphics[width=\textwidth]{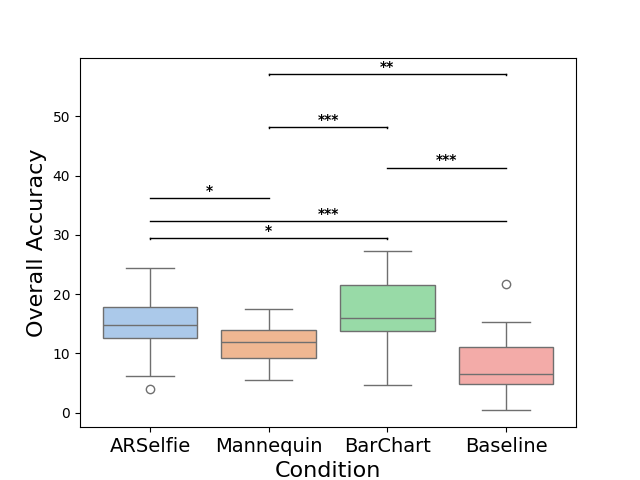}
        \caption{Activation accuracy}
        \label{fig:overall_acc}
    \end{subfigure}
    \hfill
    \begin{subfigure}[b]{0.3\textwidth}
        \centering
        \includegraphics[width=\textwidth]{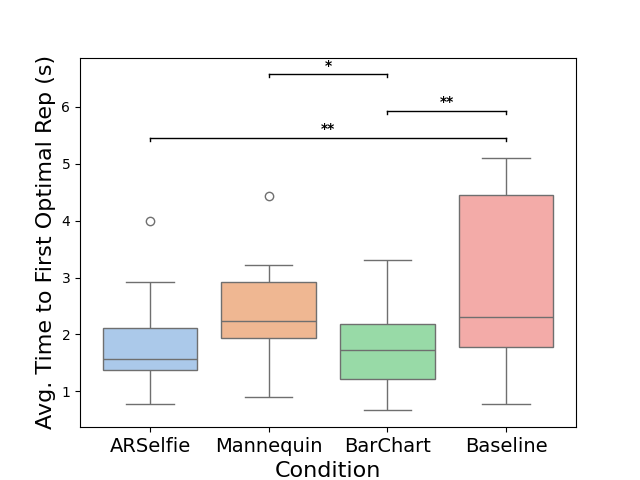}
        \caption{Time to first optimal repetition}
        \label{fig:avg_tfo}
    \end{subfigure}
    \hfill
    \begin{subfigure}[b]{0.3\textwidth}
        \centering
        \includegraphics[width=\textwidth]{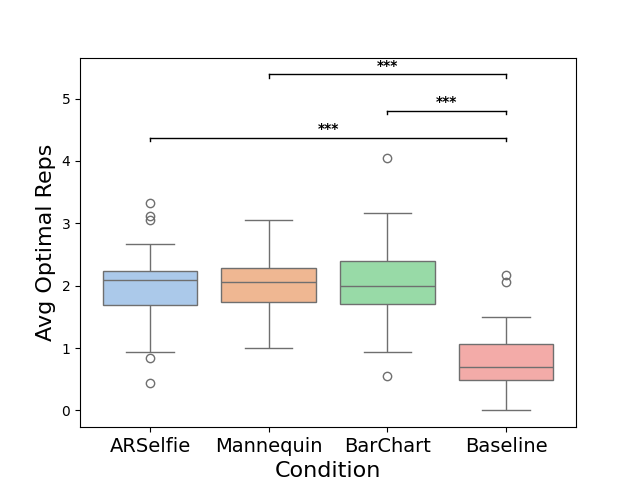}
        \caption{Number of repetitions}
        \label{fig:avg_reps}
    \end{subfigure}
    
    \caption{The performance of the participants is measured using (a) average accuracy, (b) time to first optimal repetition and (c) number of optimal repetitions. Participants were significantly more accurate with the \textit{BarChart} visualization.}
    \label{fig:perf_per_viz}
    
\end{figure}

We measured the exercise performance using three objective metrics : 
\begin{enumerate}
    \item \textbf{Accuracy}, the proportion of total exercise time the muscles were optimally activated, averaged across all exercise windows.
    \item \textbf{Number of optimal repetitions}, the average number of optimal repetitions completed within each 10-second exercise window. 
    \item \textbf{Time to first optimal repetition}, the time in seconds from the start of the exercise until the first optimal repetition was completed.
\end{enumerate}  

\add{The average performance metrics for each visualization condition are presented in Figure~\ref{fig:perf_per_viz}.}

\paragraph{Accuracy}
We found a significant main effect of the visualization condition on the average accuracy, $F_{3, 69} = 26.16, p < 0.001$. Post-hoc comparisons revealed that the participants performed significantly better in the \textit{BarChart} ($\mu = 16.88, \sigma = 5.7$) condition compared to \textit{ARSelfie} ($\mu = 14.59, \sigma = 5.3, p = 0.046$), \textit{Mannequin} ($\mu = 11.49, \sigma = 3.18, p < 0.001$) and \textit{Baseline} ($\mu = 7.97, \sigma = 5.1, p < 0.001$) conditions. Interestingly, although the \textit{Mannequin} condition is conceptually similar to the \textit{ARSelfie}, accuracy was significantly lower in the \textit{Mannequin} condition ($\mu = 11.49$) compared to \textit{ARSelfie} ($\mu = 14.59, p = 0.019$). 

\paragraph{Number of optimal repetitions}
A repeated-measures ANOVA showed significant main effect of the visualization on the average number of optimal repetitions, $F_{3,69} = 35.9, p < 0.001$. Post-hoc analysis revealed that the participants performed significantly more repetitions in the \textit{ARSelfie} ($\mu=1.97$), \textit{Mannequin} ($\mu=2.01$), and \textit{BarChart} ($\mu=2.08$) conditions than in the \textit{Baseline} condition ($\mu = 0.81$; all $p < 0.001$).

\paragraph{Time to first optimal repetition}
A Friedman test showed a significant main effect of visualization on the average time to first optimal repetition, $Q = 13.1, p = 0.004, W = 0.18$. Post-hoc pairwise comparisons indicated that participants took significantly more time to make their first optimal repetition in the \textit{Baseline} condition ($\mu = 2.85s, \sigma = 1.46s$) when compared to the \textit{ARSelfie} ($\mu = 1.76s, \sigma = 0.7s, p = 0.003$) and \textit{BarChart} ($\mu = 1.7s, \sigma = 0.65, p = 0.005$) conditions. Notably, users were also significantly slower in the \textit{Mannequin} condition ($\mu = 2.32s, \sigma=0.78s$) when compared to the \textit{BarChart} ($p = 0.01$).

\subsubsection{Learning and Carryover Effects}
We analyzed the performance data to determine if participant performance improved  as they completed more trials or if the (counterbalanced) order of conditions influenced the results. 

\begin{figure*}[!t]
    \centering
    \himg{
    \begin{subfigure}[b]{0.3\textwidth}
        \centering
        \includegraphics[width=\textwidth]{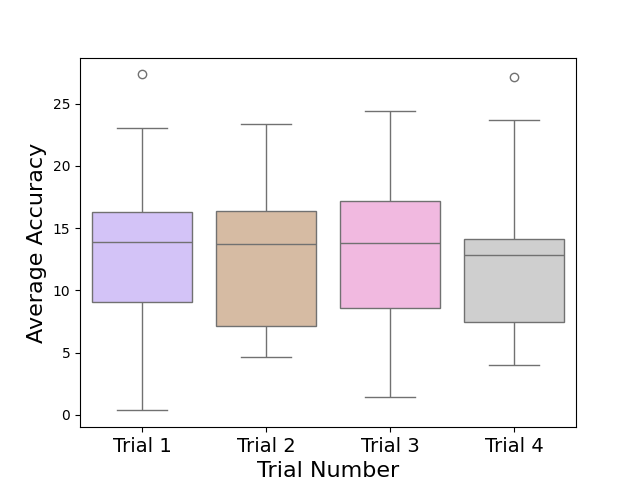}
        \caption{Activation accuracy}
        \label{fig:trial_overall_acc}
    \end{subfigure}
    \hfill
    \begin{subfigure}[b]{0.3\textwidth}
        \centering
        \includegraphics[width=\textwidth]{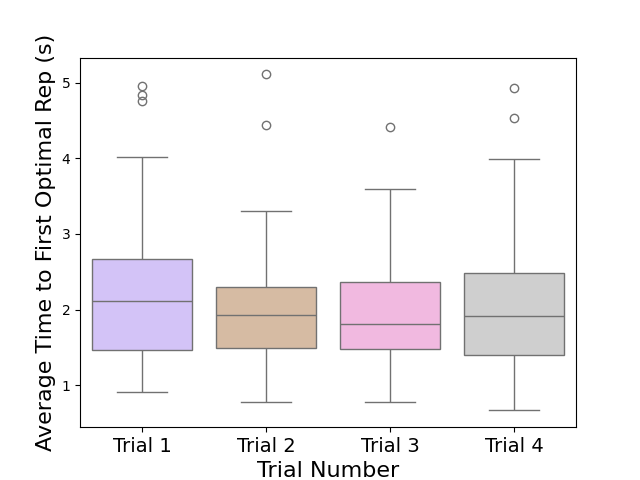}
        \caption{Time to first optimal repetition}
        \label{fig:trial_avg_tfo}
    \end{subfigure}
    \hfill
    \begin{subfigure}[b]{0.3\textwidth}
        \centering
        \includegraphics[width=\textwidth]{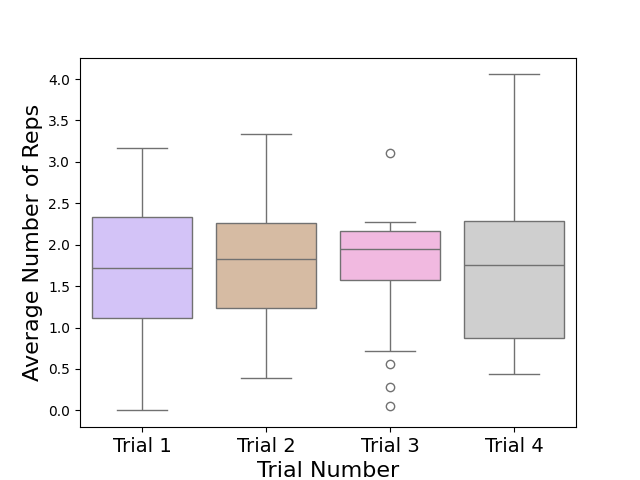}
        \caption{Number of repetitions}
        \label{fig:trial_avg_reps}
    \end{subfigure}
    \caption{The three performance metrics shown by trial number (1-4) to assess for learning effects.  The plots show that participant performance on all metrics remained stable from the first trial to the last, indicating that no significant learning effects occurred during the experiment.}
    \label{fig:trial_perf_per_viz}
    }
\end{figure*}

To test for learning effects, we assessed performance across the four trials \add{(shown in Figure~\ref{fig:trial_perf_per_viz})}. A repeated-measures ANOVA test show no significant effect of the trial number on either the average accuracy, $F_{3,69} = 0.19, p = 0.9$ or average number of optimal repetitions, $F_{3,69} = 0.04, p=0.98$. Similarly, a Friedman test revealed no significant effect of the trial number on the average time to first optimal repetition, $Q = 2.2, p = 0.53, W = 0.03$.  

\begin{figure*}[!b]
\centering
    \himg{
    \begin{subfigure}[b]{0.3\textwidth}
        \centering
        \includegraphics[width=\textwidth]{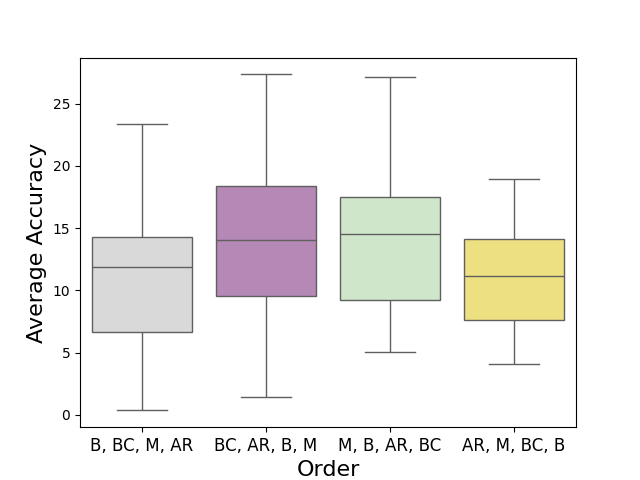}
        \caption{Activation accuracy}
        \label{fig:order_overall_acc}
    \end{subfigure}
    \hfill
    \begin{subfigure}[b]{0.3\textwidth}
        \centering
        \includegraphics[width=\textwidth]{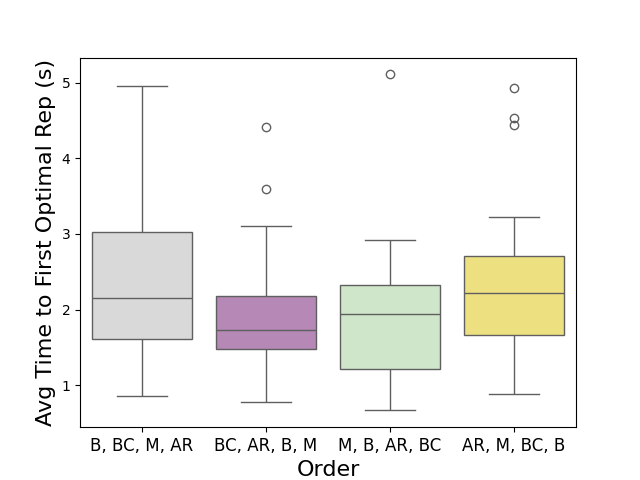}
        \caption{Time to first optimal repetition}
        \label{fig:order_avg_tfo}
    \end{subfigure}
    \hfill
    \begin{subfigure}[b]{0.3\textwidth}
        \centering
        \includegraphics[width=\textwidth]{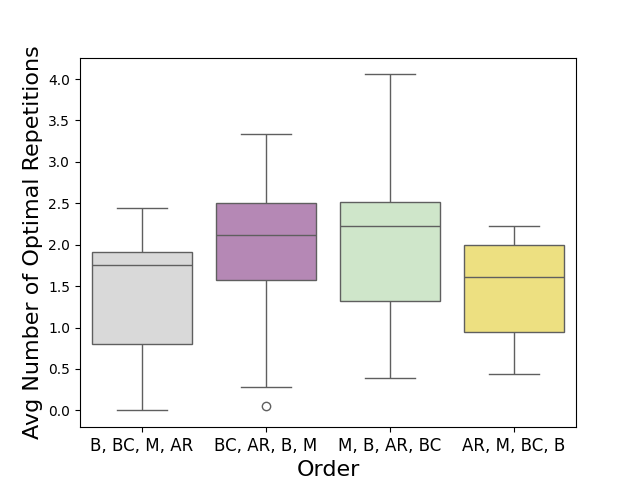}
        \caption{Number of repetitions}
        \label{fig:order_avg_reps}
    \end{subfigure}
    
    \caption{Performance metrics for each of the four presentation order groups to assess for carryover effects.  There were no significant differences in performance between the groups for any of the three metrics. This suggests that the order in which participants viewed the visualizations did not significantly influence their performance. }
    \label{fig:order_perf_per_viz}
    }
\end{figure*}

To test for carryover effects, we used a mixed-ANOVA test with condition as within-subject factor and the visualization presentation order as the between subject factor \add{(shown in Figure~\ref{fig:order_perf_per_viz})}. The analysis showed no significant effect on accuracy ($F = 1.55, p = 0.23, \eta_{p}^2 = 0.19$) or number of optimal repetitions ($F = 2.59, p = 0.08, \eta_{p}^{2} = 0.28$). For non-normal distributions we used an Aligned Rank Transform ANOVA~\cite{wobbrock2011aligned}, which also revealed no significant effect of the order, $F= 2.22, p =0.09, \eta_{p}^{2} = 0.07$. This indicates carryover effects being negligible. 

\subsubsection{Effect of ATI and Familiarity with AR}
We conducted a correlation analysis to determine if the individual differences in ATI~\cite{franke2019personal} scores or familiarity with AR influenced the performance. The analysis revealed no significant correlations between either the ATI scores or AR familiarity and any of the three performance metrics.

\subsection{User Experience}
\begin{figure}[!t]
    \centering
    
    \begin{subfigure}[b]{0.3\textwidth}
        \centering
        \includegraphics[width=\textwidth]{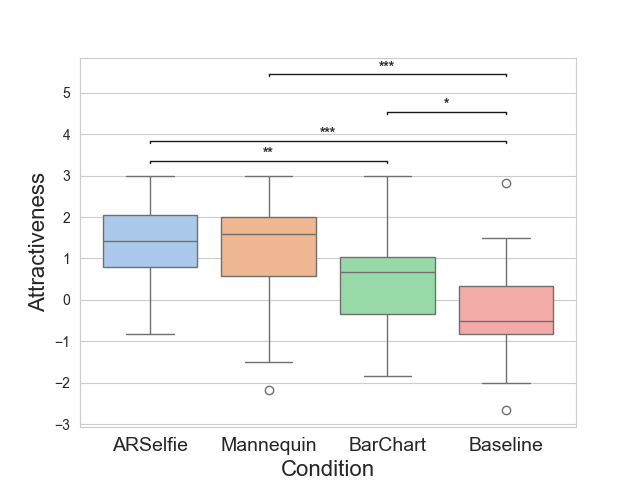}
        \caption{Attractiveness}
        \label{fig:ueq_att}
    \end{subfigure}
    \hfill
    \begin{subfigure}[b]{0.3\textwidth}
        \centering
        \includegraphics[width=\textwidth]{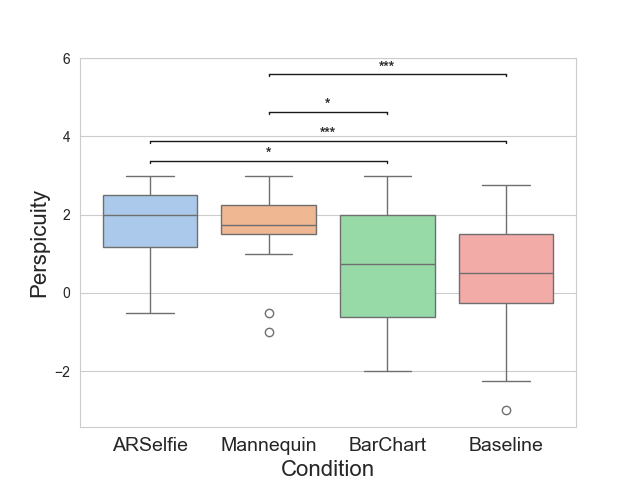}
        \caption{Perspicuity}
        \label{fig:ueq_per}
    \end{subfigure}
    \hfill
    \begin{subfigure}[b]{0.3\textwidth}
        \centering
        \includegraphics[width=\textwidth]{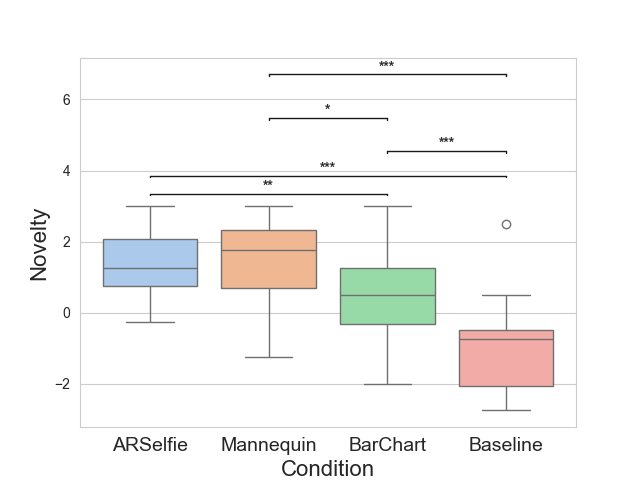}
        \caption{Novelty}
        \label{fig:ueq_nov}
    \end{subfigure}
    
    \vspace{0.5cm} 
    
    \begin{subfigure}[b]{0.3\textwidth}
        \centering
        \includegraphics[width=\textwidth]{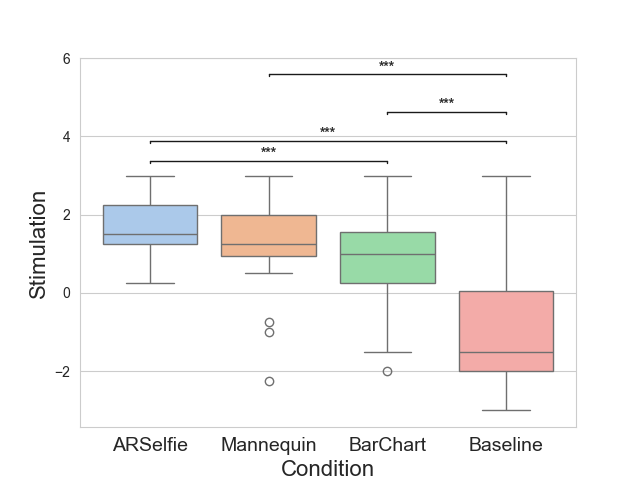}
        \caption{Stimulation}
        \label{fig:ueq_sti}
    \end{subfigure}
    \hfill
    \begin{subfigure}[b]{0.3\textwidth}
        \centering
        \includegraphics[width=\textwidth]{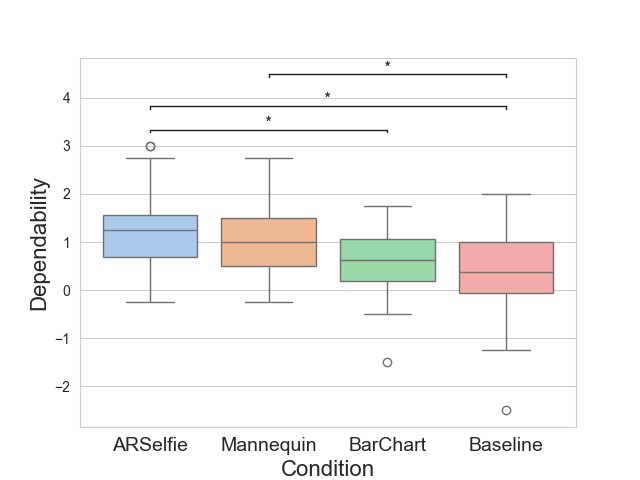}
        \caption{Dependability}
        \label{fig:ueq_dep}
    \end{subfigure}
    \hfill
    \begin{subfigure}[b]{0.3\textwidth}
        \centering
        \includegraphics[width=\textwidth]{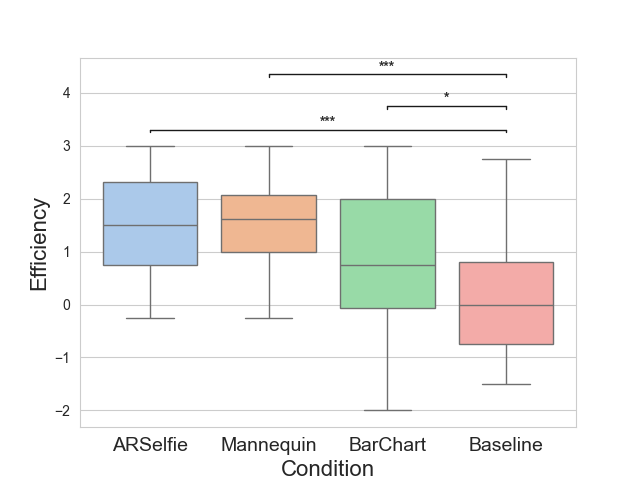}
        \caption{Efficiency}
        \label{fig:ueq_eff}
    \end{subfigure}
    
    \caption{The six sub-scales of the User Experience Questionnaire~\cite{schrepp2017construction} for each of the experimental conditions. The whiskers indicate significant post-hoc comparisons (* p < 0.05; ** p < 0.01, *** p < 0.001). Higher is better with scores > 0.8 considered positive and < -0.8 considered negative. }
    \label{fig:ueq}
    
\end{figure}

The user ratings across the six scales of the User Experience Questionnaire (UEQ)~\cite{schrepp2017construction} for the four conditions is shown in Figure~\ref{fig:ueq}. Overall, the \textit{ARSelfie} and \textit{Mannequin} conditions consistently scored highest across both the pragmatic and hedonic aspects of user experience. The \textit{BarChart} condition was generally rated higher than \textit{Baseline} but lower than the two embedded conditions.

\textbf{Pragmatic quality} refers to the task-related aspects of the user experience. It is composed of the \textit{Perspicuity}, \textit{Efficiency}, and \textit{Dependability} sub-scales. 
For \textit{Perspicuity}, a Friedman test showed a significant main effect of the visualization condition $(W = 0.35, p < 0.001)$. Post-hoc comparisons indicated that the \textit{ARSelfie} and \textit{Mannequin} conditions were perceived as significantly clearer than \textit{BarChart} ($p = 0.019$ and $p = 0.016$, respectively) and \textit{Baseline} ($p < 0.001$ for both).

For \textit{Efficiency}, a Friedman test revealed a significant main effect ($W = 0.5, p < 0.001$). Post-hoc comparisons showed that the participants perceived the \textit{Baseline} condition as significantly less efficient in guiding them than \textit{ARSeflie} ($p < 0.001$), \textit{Mannequin} ($p < 0.001$), and \textit{BarChart} ($p = 0.026$). No significant differences were found among the other three conditions.

For \textit{Dependability}, a repeated measures ANOVA found a significant main effect ($F_{3, 69} = 8.41, p < 0.001, \eta^{2}_g = 0.16$), with a Greenhouse-Geisser correction applied as the sphericity assumption was violated. Post-hoc tests showed that the participants found the \textit{ARSelfie} condition more dependable than both \textit{BarChart} ($p = 0.023$) and \textit{Baseline} ($p = 0.012$). The \textit{Mannequin} visualization was also rated as more dependable than the \textit{Baseline} condition ($p=0.011$).

\textbf{Hedonic quality} measures the affective aspects of the interaction. It consists of the \textit{Stimulation} and \textit{Novelty} sub-scales.

For \textit{Stimulation}, a Friedman test indicated a significant main effect ($W = 0.65, p < 0.001$). The \textit{Baseline} condition was rated as significantly less stimulating than the \textit{ARSelfie} ($p < 0.001$), \textit{Mannequin} ($p < 0.001$) and \textit{BarChart} ($p < 0.001$). Moreover, the \textit{ARSelfie} was rated more stimulating than \textit{BarChart} ($p < 0.001$). 

For \textit{Novelty}, a Friedman test revealed a significant main effect ($W = 0. 62, p < 0.001$). Post-hoc comparisons indicate a definite ordering, with \textit{ARSelfie} and \textit{Mannequin} rated as more novel, followed by \textit{BarChart} with \textit{Baseline} being rated as least novel. All the pairwise comparisons along this ordering except the one between \textit{ARSelfie} and \textit{Mannequin} were significant ($p < 0.05$). 

The \textit{Attractiveness} scale provides a global measure of each condition's overall appeal. A repeated measures ANOVA found a significant main effect ($F_{3, 69} = 15.63, p < 0.001, \eta_{g}^{2} = 0.24$). The \textit{Baseline} condition was rated as significantly less attractive than \textit{ARSelfie} ($p < 0.001$), \textit{Mannequin} ($p < 0.001$) and \textit{BarChart} ($p = 0.014$). Additionally, \textit{ARSelfie} was rated as more attractive than \textit{BarChart} ($p = 0.008$).

\subsection{Task Load}
\begin{figure*}
    \centering
    \begin{subfigure}[b]{0.45\textwidth}
    \centering \includegraphics[width=\linewidth]{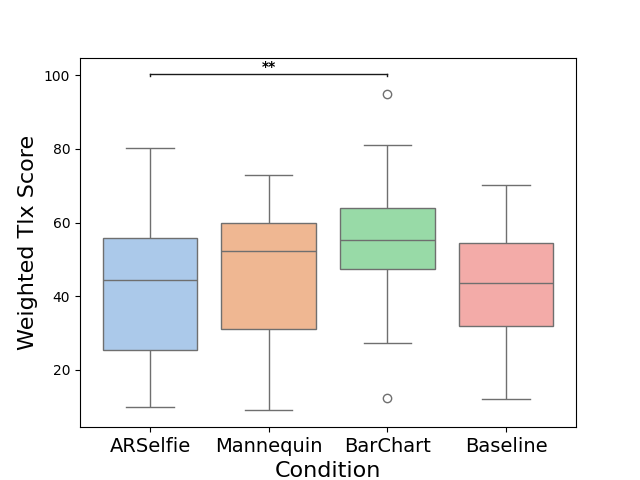}
    \caption{Weighted NASA-TLX Score.}
        \label{fig:nasa}
    \end{subfigure}
    \hfill
    \begin{subfigure}[b]{0.45\textwidth}
        \centering
        \includegraphics[width=\linewidth]{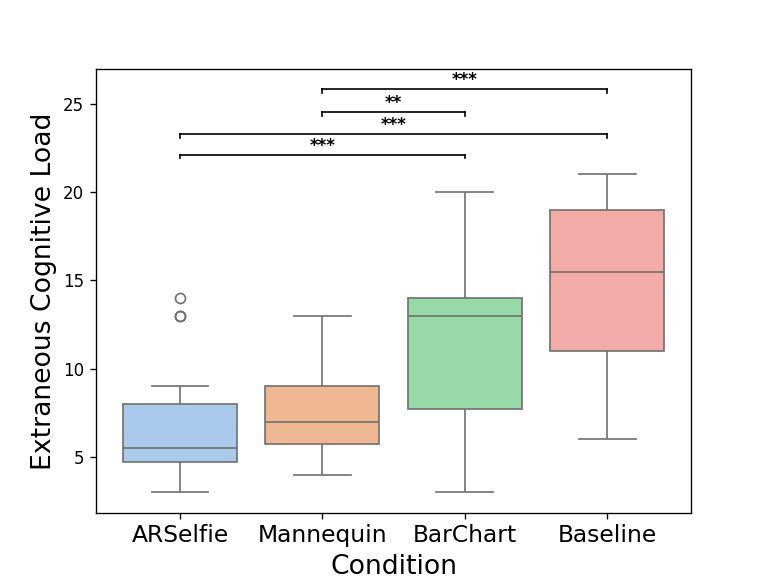}
        \caption{Extraneous Cognitive Load}
        \label{fig:ecl}
    \end{subfigure}
    
    \caption{(a) The weighted NASA-TLX scores for each of the conditions (lower is better). Post-hoc comparisons showed a significantly higher task load in the \textit{BarChart} condition when compared to \textit{ARSelfie}. \add{(b) The ECL scores for the four visualization conditions (lower is better). The ECL associated with embedded conditions were significantly lower than the \textit{BarChart} and \textit{Baseline} conditions.}}
    \label{fig:loads}
    
\end{figure*}

A repeated-measures ANOVA was conducted to assess the impact of the visualization condition on the weighted overall NASA-TLX workload score. As the data violated the assumption of sphericity a Greenhouse-Geisser correction was applied. The analysis revealed a significant main effect, $F_{3, 69} = 4.6, p = 0.011, \eta_g^{2} = 0.083$. Post-hoc pairwise comparisons using Bonferroni-Holm corrections showed that the workload for the \textit{ARSelfie} condition (\add{$\mu = 42.8, \sigma=18.6$}) was significantly lower than the \textit{BarChart} condition ($\add{\mu=55.7, \sigma=17.6}, p = 0.009$). \del{Interestingly,} The overall workload in the \textit{Baseline} condition ($\mu = 43.19, \sigma=16.1$) was \del{lesser} lower than in the \textit{BarChart} (\add{$\mu=55.7, \sigma=17.6$}) and \textit{Mannequin} ($\mu = 46.5, \sigma=17.9$) conditions. However, the \add{pairwise} differences were not significant. The results are shown in Figure~\ref{fig:nasa}.

\begin{figure*}
    \centering
    
    \begin{subfigure}[b]{0.3\textwidth}
        \centering
        \includegraphics[width=\textwidth]{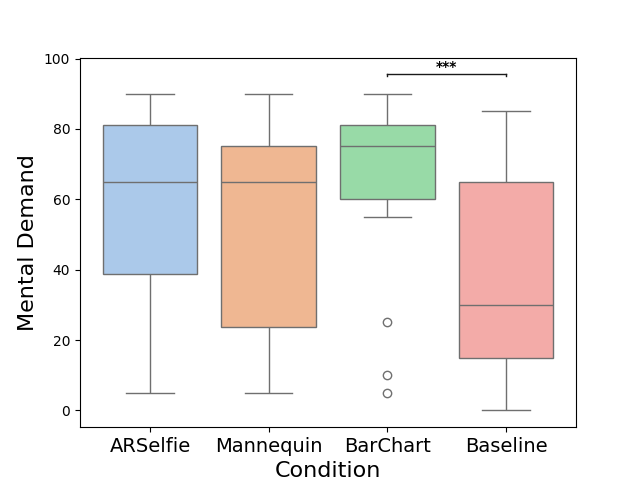}
        \caption{Mental Demand}
        \label{fig:tlx_md}
    \end{subfigure}
    \hfill
    \begin{subfigure}[b]{0.3\textwidth}
        \centering
        \includegraphics[width=\textwidth]{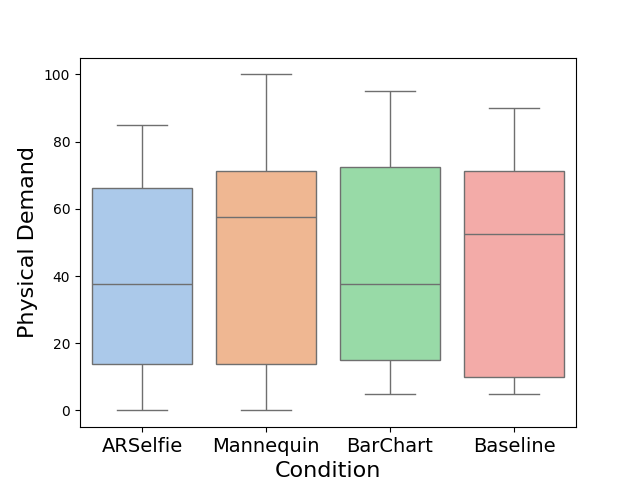}
        \caption{Physical Demand}
        \label{fig:tlx_pd}
    \end{subfigure}
    \hfill
    \begin{subfigure}[b]{0.3\textwidth}
        \centering
        \includegraphics[width=\textwidth]{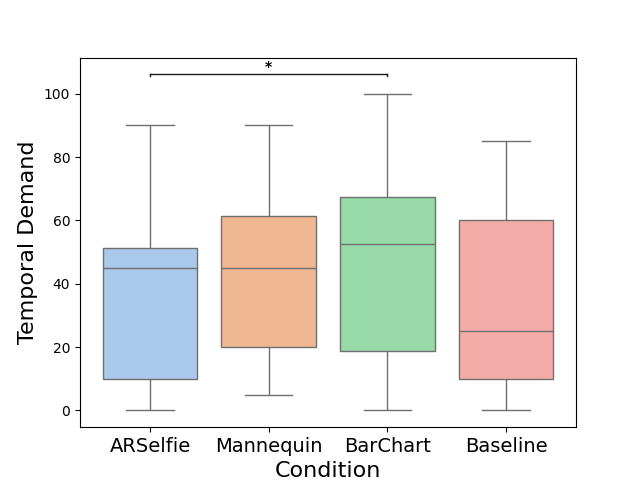}
        \caption{Temporal Demand}
        \label{fig:tlx_td}
    \end{subfigure}
    
    \vspace{0.5cm} 
    
    \begin{subfigure}[b]{0.3\textwidth}
        \centering
        \includegraphics[width=\textwidth]{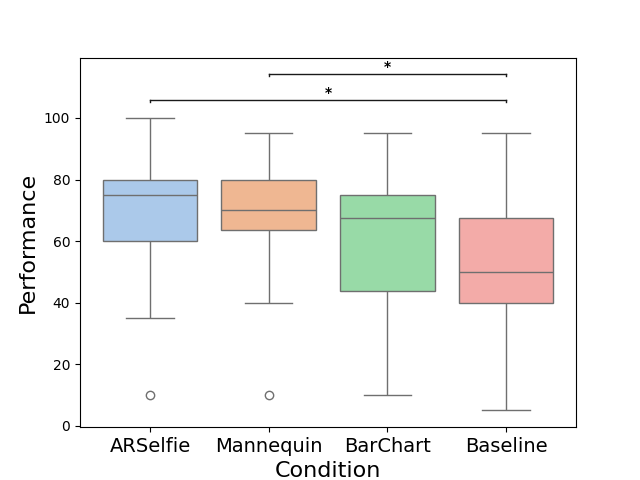}
        \caption{Performance}
        \label{fig:tlx_perf}
    \end{subfigure}
    \hfill
    \begin{subfigure}[b]{0.3\textwidth}
        \centering
        \includegraphics[width=\textwidth]{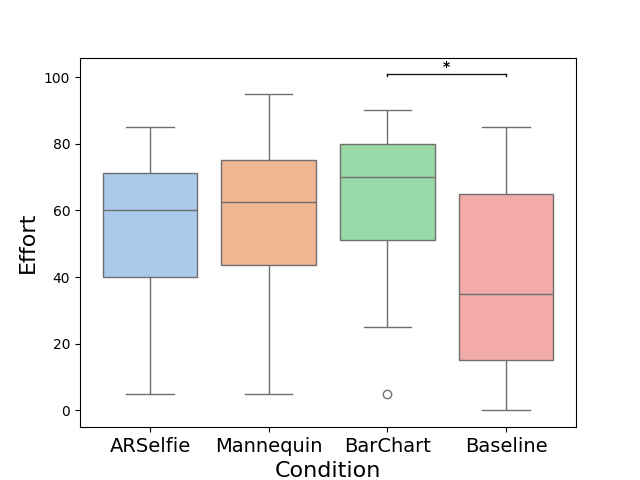}
        \caption{Effort}
        \label{fig:sub5}
    \end{subfigure}
    \hfill
    \begin{subfigure}[b]{0.3\textwidth}
        \centering
        \includegraphics[width=\textwidth]{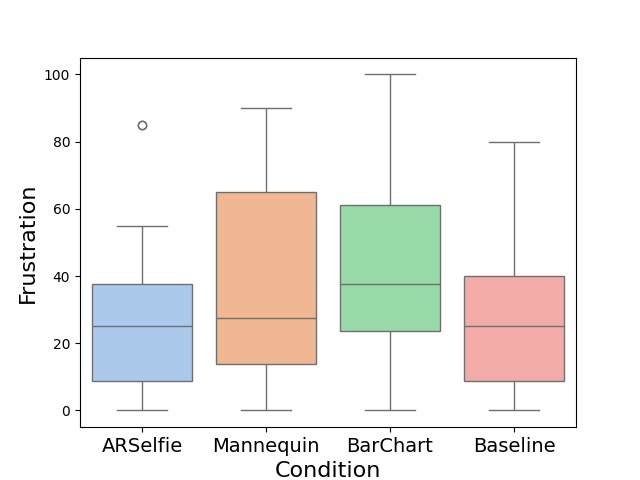}
        \caption{Frustration}
        \label{fig:sub6}
    \end{subfigure}
    
    \caption{The NASA-TLX sub-scale scores for each of the visualization conditions. Lower scores are better for all scales except \textit{Performance}. The whiskers indicate significant post-hoc differences (* p < 0.05; ** p < 0.01, *** p < 0.001).}
    \label{fig:nasa_sub_scales}
\end{figure*}

To further investigate the sources of task load, we analyzed the six individual TLX sub-scales. The results are shown in Figure~\ref{fig:nasa_sub_scales}.

\subsection{Extraneous Cognitive Load}
Extraneous cognitive load (ECL) is caused by the design of the learning material or application~\cite{klepsch_development_2017, sweller1994cognitive}. Users often have to invest cognitive resources in extraneous processes such as searching for information, ignoring irrelevant information while learning or performing a task which might impair their efficiency~\cite{sweller1994cognitive}. 
We assessed ECL using a modified version of the questionnaire used in ~\citet{klepsch_development_2017}. It consists of three 7-point Likert-scale questions (see Table ~\ref{tab:ecl_questions}) with Cronbach's $\alpha = 0.82$. The overall ECL is computed as the sum of the three item scores. The results are shown in Figure~\ref{fig:ecl}.

\begin{table}[h!]
    \centering
     \caption{The questionnaire used to assess extraneous cognitive load. Responses were captured on a 7-point scale (1=\textit{Strongly Disagree}, 5=\textit{Strongly Agree}). The items were adapted from the ~\cite{klepsch_development_2017} to assess how hard it was for participants to obtain the information needed from the visualization to activate their muscles optimally.}
    \begin{tabularx}{0.99\linewidth}{l X}
        \toprule
         1. & It was exhausting to find the muscle activation information needed from this visualization \\
         2. & This visualization is inconvenient for verifying if the muscles are activated optimally \\
         3. & This visualization made it difficult to recognize information and link it to the task \\
         \bottomrule
    \end{tabularx}
   
    \label{tab:ecl_questions}
\end{table}

A Friedman test revealed a significant main effect of visualization condition on the ECL, $(Q = 21.78, W = 0.49, p < 0.001)$. Post-hoc comparisons showed that participants reported significantly lower ECL in the \textit{ARSelfie} ($\mu = 6.50, \sigma = 3.16$) and \textit{Mannequin} ($\mu = 7.71, \sigma = 2.53$) conditions compared to both the \textit{BarChart} ($\mu = 11.42, \sigma = 4.77$; $p = 0.001$ and $p = 0.008$, respectively) and \textit{Baseline} ($\mu = 14.88, \sigma = 4.79$; $p < 0.001$ for both). Additionally, \textit{Mannequin} was rated as imposing significantly less ECL than \textit{BarChart} ($p = 0.008$). No significant difference was observed between the embedded conditions.

\section{Qualitative Results}

\subsection{Open-ended questions}
After each visualization condition, participants provided written responses to the question, \textit{``What did you like about this visualization}.'' After the whole study they provided written responses to the question, \textit{``Which visualization did you prefer overall? Why?}.'' Finally, they provided open-ended feedback about the different visualization conditions in a semi-structured interview with the researcher. Participants were specifically asked to elaborate on the aspects of the visualizations that they liked and why they preferred one condition over the other. Their responses to these open-ended questions were coded and organized into themes using inductive thematic analysis~\cite{braun2006using}. 

\subsubsection*{\textbf{Theme 1: Cognitive and Task Load}}

Participants frequently noted that both the location of the feedback and its visual encoding affected the amount of effort required to incorporate it. 

\paragraph{Visualization Location}

The \textit{BarChart} condition required frequent gaze shifts between the bars and the face which many found effortful, with P6 mentioning, ``looking down and up made it harder to assess the state'' and P23 was ``annoyed a little bit about having to look up and down''. Others found these gaze shifts distracting, such as P10 who ``could not concentrate on the face and kept looking at the feedback'' and P11 who ``liked it (\textit{BarChart} condition) the least as I had to look in two different places''. On the other hand, the co-located feedback in the embedded conditions reduced gaze shifts. P24 in the \textit{ARSelfie} condition mentioned that they, ``didn't have to  switch [their] attention to different parts of the screen'' and P7 in the \textit{Mannequin} condition highlighted they only needed to ``pay attention to one region.'' 

\paragraph{Visual Encoding}

Participants found the proportional bar fills with color changes in the \textit{BarChart} condition provided clear, quantifiable feedback. Three out of 24 participants (P14, P22, P23) noted that the \textit{BarChart} showed them ``how much [they] had to move''. The opacity-based encodings in the embedded conditions were considered less-intuitive. While P5 found them useful for tracking activation five out of 24 participants (P8, P14, P18, P22, P26) described them as more difficult to interpret, requiring more concentration. P18 explained that it was ``more effortful to concentrate on the opacity [and] easier to focus on the left/right of the bar chart'' while P14 summarized that ``transparency was a lot harder to interpret than the bar fill.''  

\subsubsection*{\textbf{Theme 2: Perceived Performance and Learning}}
    Participants pointed out different strengths of \del{the} each visualization. 

\paragraph{Learning}
Seven out of 24 participants (P7, P14, P17, P18, P19, P24, P26) perceived the embedded conditions as better for learning how to control the facial muscles. P7 found that \textit{ARSelfie} condition made ``it a little easier to learn how much force I need to apply over time''. Both P19 and P24 mentioned that seeing their own facial form in the \textit{ARSelfie} condition made it easier to remember correct movements. Even though P18 preferred another visualization overall (\textit{BarChart}), they stated that the \textit{Mannequin} condition ``made it very easy to learn which muscle needed to move in what way'' while P26 thought that the \textit{Mannequin} condition had the ''clearest approach to showing muscle movement that I could learn and adjust from.'' 

In contrast, participants reported limited learning in the \textit{BarChart} condition, because either they often ignored the muscle labels or could not map them to regions of their face. P17 admitted that they ``didn't even read the labels and just looked at the colors'' and they ``[were] not learning at all''. Similarly, P24 mentioned that they ``didn't read the labels'' so they ``could not figure out the correspondence [between the feedback and facial muscles]''.  Some participants despite reading the muscle labels struggled with the mapping. P3 mentioned being ``not able to figure [out] what particular part had to be changed'' and P26 similarly stated that ``it was hard to correlate the muscle names to specific regions of the face''. Even when the mapping was understood, the additional effort required for the attention shifts reduced the utility of the feedback and hindered learning. P6 recalled that ``mapping feedback to movement took time and brain power,'' which may have led to some participants (e.g., P21) to stop observing their face and focus only on the feedback. 

\paragraph{Performance}
Many felt that the \textit{BarChart} condition was effective for immediate performance. P18 noted that the proportional bar fills eliminated ``ambiguity from the color/opacity system'' and enabled ``to tweak things much more easily''. P14 stated they ``liked being able to see how close to optimal activation'' they were and observing ``how much [their] muscle movements changed the score''. On the other hand, the feedback overlays in the embedded conditions apparently made adjustment harder for some participants. P21 mentioned that the ``filters made it slightly harder to know'' if their muscle movements were correct and P22 voiced that they ``found it hard to correlate what my face was like because of the overlay''

\subsubsection*{\textbf{Theme 3 : User Experience and Preference}}
Many participants preferred the \textit{ARSelfie} condition as they found it easier to relate their movements with their facial form (P3, P5, P19, P20) or simply because they preferred looking at their own face (P23, P24) instead of the avatar.  But several (P4, P6, P21, P26) participants reported discomfort, either from self-consciousness or distractions. P21 mentioned they ``didn't like looking at [their] face and ... ended up focusing on the features'' and P4 stated that it was ``distracting to look at the face and did not want to look at the face for long''. 

For \del{a} some of the participants the \textit{Mannequin} condition served as a distraction-free alternative to the \textit{ARSelfie}. P4 mentioned that ``it eliminated the background and made it easier to just focus on the facial expressions''. P6, P9, P21 all expressed that it had ``no distractions'' and made it easier to focus on the exercises. However, a few participants found the \textit{Mannequin} condition ``unnatural'' (P11, P19), ``less interactive'' (P23) and even ``creepy'' (P20).

\subsection{Preferences}
\begin{figure}
    \centering
    \includegraphics[width=0.5\linewidth]{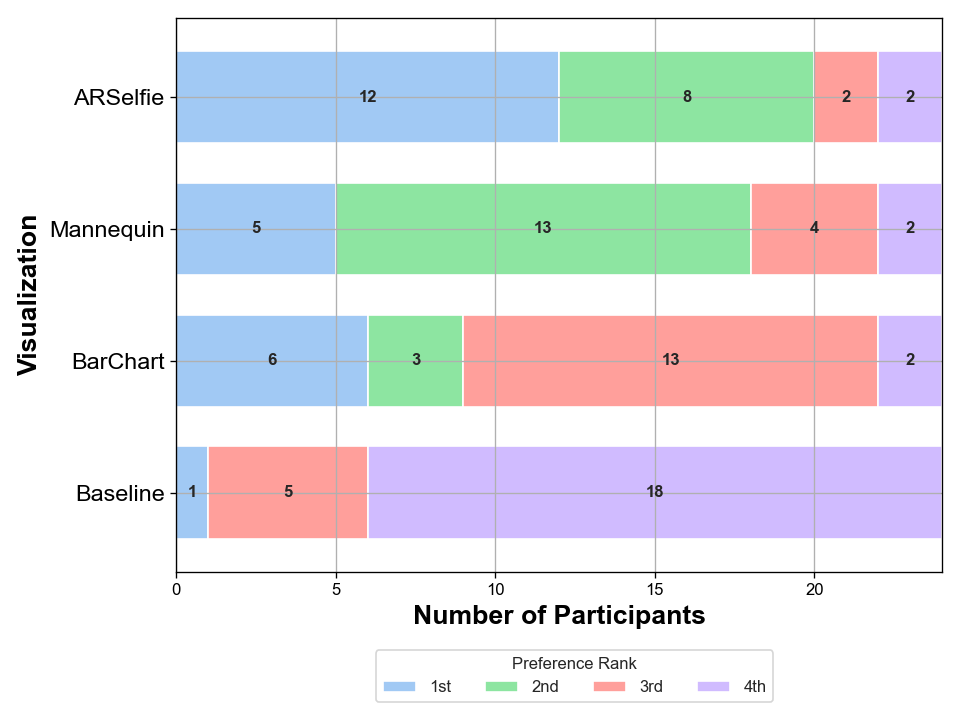}
    \caption{User preference rankings for the four visualization conditions. \textit{ARSelfie} condition received the highest number of first-place rankings, while the \textit{Mannequin} was most frequently ranked second. \textit{BarChart} was most often ranked third, and Baseline was the least preferred condition, receiving the highest number of fourth-place rankings.}
    \label{fig:prefs}
\end{figure}

The aggregate rankings (shown in Figure~\ref{fig:prefs}) from most preferred to least preferred were: \textit{ARSelfie}, \textit{Mannequin}, \textit{BarChart} and \textit{Baseline}. The \textit{ARSelfie} condition was the most preferred, with 20 out of 24 participants ranking it in their top two, followed by the \textit{Mannequin} condition (18 of 24). In contrast, the \textit{BarChart} (9 of 24) and \textit{Baseline} (6 of 24) conditions were ranked in the top two much less frequently. A Friedman test confirmed that the visualization condition had statistically significant effect on preference rankings, $Q = 29.75, p < 0.001$ with a moderate-to-strong effect size (Kendall's $W = 0.41$). Post-hoc pairwise comparisons using Bonferroni corrections showed that the \textit{Baseline} condition was ranked significantly lower than the \textit{ARSelfie}, \textit{Mannequin} and \textit{BarChart} conditions. No other pairwise comparisons were statistically significant. 

\section{Discussion and Design Implications}
 Our findings show that spatial placement of feedback strongly shapes workload, user experience and performance in facial exercises. We interpret these results to derive strategies for designing future facial exercise systems. 

\subsection{Embedded Feedback for the Face?}
Our quantitative and qualitative results indicate that the benefits of embedded feedback, previously demonstrated for whole-body movements, extend to facial movements as well. \add{NASA-TLX and ECL results indicate that the spatial alignment of visual cues with the region of movement in the embedded conditions (\textit{ARSelfie} and \textit{Mannequin}) enabled users to directly associate feedback with the target facial muscles. It is plausible that this would minimize the need for spatial translation~\cite{larsen1998effects} between the feedback and the targets, resulting in lower task and cognitive load compared to the \textit{BarChart} and \textit{Baseline} conditions (\textbf{RQ2}).} \del{The spatial alignment of visual cues with the region of movement enabled users to directly associate feedback with the target facial muscles. This reduced the cognitive demands of spatial translation~\cite{larsen1998effects} between the feedback and the targets.} 
In contrast, the \textit{BarChart} condition was cognitively demanding as it required users to mentally translate feedback presented in a peripheral location onto their face to make adjustments. This spatial separation likely increased \textit{cognitive distance}~\cite{kim2009simulated} between the feedback and its referent, increasing the chances of the \textit{split-attention} effect~\cite{chandler1992split}.  
The significantly higher score on the NASA-TLX temporal demand sub-scale for the \textit{BarChart} condition, compared to the \textit{ARSelfie}, further suggests that attention shifts between the face and feedback combined with the spatial translation process contributed to participants feeling more rushed. Although participants were more accurate in activating their muscles while using the \textit{BarChart} condition, the performance gain likely came at the cost of learning and cognitive effort.
The higher perspicuity and efficiency scores for the embedded visualizations further indicate that spatially congruent feedback was easier to integrate. This aligns with the \textit{spatial continuity principle}~\cite{mayer201412}. 

Designers of facial exercise feedback systems should leverage spatially congruent visualizations that align directly with the target muscles as they reduce extraneous cognitive and task load imposed on the user \add{(\textbf{RQ2})}. While situated feedback modalities support accurate movements \add{(\textbf{RQ1})}, designers must be mindful of the cognitive demands associated with the attention shifts and spatial translation. Broadly, these findings indicate that embedded feedback design principles, established for whole-body or upper-limb training can be extended to facial exercises.   

 \subsection{Selfie Tradeoffs}
 \add{The embedded conditions were generally preferred to the \textit{BarChart} and \textit{Baseline} conditions \textbf{(RQ3)} with the} \textit{ARSelfie} condition \del{was}rated highest in terms of user experience and \del{was} the most preferred overall, \del{which} indicating the effectiveness of spatially aligned feedback for facial movement tasks. However, qualitative feedback revealed important tradeoffs for designing embedded feedback. Even though the participants valued the spatial benefits of the embedded conditions, their opinions on the selfie-view varied considerably. Some found visual overlays on their face interactive and intuitive, but for others seeing their face was distracting and uncomfortable. 

 These contrasting responses align with research on self-focused attention, which suggests that individuals are hardwired to attend to their faces and this can override top-down attentional control and lead to cognitive exhaustion~\cite{wojcik2018self, bredart2006short}. This could be compounded by self-evaluation, mirror anxiety and ``Zoom fatigue,'' where sustained focus on one's own image can trigger negative self-perception and deplete attentional resources~\cite{bailenson2021nonverbal, carolli2023zoom}. In this state, a user's attention may be captured by their own perceived flaws, rendering them effectively blind to the AR feedback itself~\cite{kreitz2015inattentional}. These concerns might have partially offset the benefits of spatial alignment in some participants.

 The \textit{Mannequin} condition mitigated this by providing a less personal proxy. While this approach did reduce self-consciousness for some, others described the avatar as ``unnatural'' or ``creepy'', indicating sensitivity to uncanny valley effects~\cite{mori2012uncanny}.
 
Our findings surface a core design trade-off. While overlaying feedback on a face, whether the user's own or an avatar's, might provide superior spatial guidance, its effectiveness can be enhanced by accommodating individual differences in self-representation comfort.

We recommend that embedded facial feedback systems incorporate flexible self-representation options to maximize the spatial benefits while minimizing psychological barriers. This could include progressive abstraction levels (from stylized avatars to photorealistic), customizable avatar features to reduce uncanny valley effects, using proxies of favorite fictional characters, or hybrid approaches that combine abstract overlays with minimal facial landmarks. This could allow users to fully benefit from the advantages of embedded feedback. 

\subsection{Perceptual Clarity}

While our results indicate that the benefits of embedded feedback extend to the face, their effectiveness is highly dependent on the perceptual clarity of the cues. 
 
First, participants struggled to interpret the magnitude of corrective adjustment conveyed through opacity modulation. This encoding limitation created a preference-performance trade-off, while users preferred embedded conditions for its spatial advantages, they activated their muscles more accurately leveraging the explicit magnitude representation in the \textit{BarChart} condition. Second, the AR overlay itself introduced some visual clutter. Some participants found that the semi-transparent face filters obscured the view of the facial muscles they were trying to control. This visual interference can be detrimental for learning, as clear visual feedback is  critical for developing accurate motor control especially during the skill acquisition stage~\cite{robin2004sensory}. 

Therefore, we recommend designers combine spatially embedded feedback with explicit magnitude indicators. While spatial cues are effective for showing where the users should focus, they should be supplemented with perceptually effective cues such as proportional fill of the muscle area or outlines or numerical percentages in text to clearly communicate how much adjustment is needed to reach the target. These filters must be designed carefully to provide feedback without obscuring the user's self view. 

\subsection{Feedback Geometry} 
The \textit{BarChart} condition used 2D bars with consistent shape and size for all muscles, whereas the embedded conditions used overlays that followed the natural, 3D contours of individual facial muscles. This difference could have influenced how participants perceived and interpreted the feedback. The bar fills potentially benefited from their uniform geometry as changes were visually consistent across muscles and hence, easier to compare. But it required frequent attentional shifts. 

In contrast, the overlays provided spatially aligned feedback that made it easier for participants to connect it with the muscles they were exercising but these overlays varied in shape and extent. The changes in larger muscles (e.g., forehead) could have been more conspicuous while the changes in small or thin muscle overlays were subtle and harder to detect. This uneven salience of feedback across muscles presents a challenge unique to facial embedded feedback, while spatial alignment reduces effort and enhances user experience, geometric variability of the overlays can create perceptual imbalances. 

When using embedded feedback in facial exercise systems designer should ensure that they normalize perceptual salience by adjusting visual parameters such as brightness or contrast to amplify feedback on smaller or less conspicuous muscles. They can also use supplementary indicators like subtle outlines or glyphs to provide consistent salience across muscles.

\section{Limitations and Future Work}

Our work indicates that spatially aligned, embedded visualizations can effectively support facial motor training by reducing cognitive load and enhancing user experience compared to situated visualizations. However, several limitations qualify our findings and point to future directions.

First, the study was conducted in a controlled laboratory setting lasting 45–60 minutes and focused on three exercises involving 11 distinct facial muscles. While this scope allowed focused comparison across visualization conditions, it leaves open questions about the long-term effectiveness of embedded feedback in supporting the learning and retraining of facial movements. Future work should \add{assess learning gains associated with each visualization condition through a between-subjects study and} explore longitudinal use, broader exercise repertoires, and more ecologically valid settings to assess sustained outcomes.

Second, our evaluation focused exclusively on visual feedback. Even within a short session, some participants reported discomfort when repeatedly viewing their own face in the \textit{ARSelfie} condition. This discomfort may be amplified in clinical or rehabilitation contexts, potentially discouraging engagement. Future systems should examine avatar-based or proxy feedback strategies in greater depth, and investigate non-visual modalities such as auditory or haptic cues—particularly in contexts where repeated face viewing is not viable or desirable.

Third, our feedback pipeline uses the ARCore face mesh with a manual mapping of vertices to underlying musculature \add{and heuristics to infer muscle activation states from detected facial movements, both} validated by a facial anatomy expert. While this mapping provides anatomical grounding, it remains an indirect proxy for muscle activation and does not capture physiological signals such as recruitment strength or neuromuscular coordination. \add {Before real-world deployment for facial muscle training applications, this ARCore-based muscle activation detection system and its underlying heuristics should be validated against ground-truth physiological measurements such as EMG to ensure accuracy and reliability.}
Future work could \add{also use sophisticated algorithms such as ~\cite{liu2025pianoemg} to predict muscle activation directly from image data,} integrate complementary sensing modalities (e.g., EMG) or incorporate biomechanical models of facial anatomy to enable more precise and potentially diagnostic feedback.

Finally, our participant sample was relatively small and homogeneous in age, which limits generalizability. Future studies should recruit larger and more diverse populations, including clinical groups and individuals with varied expressive norms or cultural backgrounds, to better assess the applicability and inclusivity of our design insights.

\section{Conclusion}

In this work, we investigated how the spatial placement of visual feedback affects performance, cognitive load, and user experience in facial motor training. We compared three visualization techniques of situated (BarChart), proxy-embedded (Mannequin), and embedded (ARSelfie) and a Baseline condition where participants saw their face in the mobile device's front-facing camera but received no additional feedback on facial muscle activation, in a within-subjects study (N=24).
Our results indicate that spatially aligned feedback reduces cognitive effort, improves user experience, and is generally preferred over situated alternatives. While situated feedback can support greater accuracy, it imposes higher cognitive demand due to attention shifts and the need for spatial translation.
Based on these findings, we provide design implications for future AR-based facial feedback systems, with implications for rehabilitation, performance training, and skill acquisition.

\bibliographystyle{ACM-Reference-Format}
\bibliography{intarface}

\end{document}